\documentclass[apjl]{emulateapj}
\usepackage{color}

\usepackage{graphicx}% http://ctan.org/pkg/graphicx

\usepackage[colorlinks=true, allcolors=blue]{hyperref}
\usepackage{lineno}
\begin{document}

%\title{The Scorpion Planet Survey: Demographics of Directly Imaged Giant Exoplanets on Wide Orbits within a Nearby Group of Young A-type Stars}
\title{The Scorpion Planet Survey: Wide-Orbit Giant Planets Around Young A-type Stars}

\color{black}
\shorttitle{Scorpion Planet Survey}
\shortauthors{Wagner et al.}
\author{Kevin Wagner,\altaffilmark{1,2,3}$^{\star}$ D\'aniel Apai,\altaffilmark{1,2,4} Markus Kasper,\altaffilmark{5}, Melissa McClure,\altaffilmark{6} and Massimo Robberto\altaffilmark{7}}

\altaffiltext{1}{Steward Observatory, University of Arizona}
\altaffiltext{2}{NASA NExSS \textit{Alien Earths} Team}
\altaffiltext{3}{NASA Hubble Fellowship Program $-$ Sagan Fellow}
\altaffiltext{4}{Lunar and Planetary Laboratory, University of Arizona}
\altaffiltext{5}{European Southern Observatory, Germany}
\altaffiltext{6}{University of Leiden, Netherlands}
\altaffiltext{7}{Space Telescope Science Institute}

\altaffiltext{$\star$}{Correspondence to: kevinwagner@email.arizona.edu}


\begin{abstract}

The first directly imaged exoplanets indicated that wide-orbit giant planets could be more common around A-type stars. However, the relatively small number of nearby A-stars has limited the precision of exoplanet demographics studies to $\gtrsim$10\%. We aim to constrain the frequency of wide-orbit giant planets around A-stars using the VLT/SPHERE extreme adaptive optics system, which enables targeting $\gtrsim$100 A-stars between 100$-$200 pc. We present the results of a survey of 84 A-stars within the nearby $\sim$5$-$17 Myr-old Sco OB2 association. The survey detected three companions$-$one of which is a new discovery (HIP75056Ab), whereas the other two (HD 95086b and HIP65426b) are now-known planets that were included without \textit{a priori} knowledge of their existence. We assessed the image sensitivity and observational biases with injection and recovery tests combined with Monte Carlo simulations to place constraints on the underlying demographics. We measure a decreasing frequency of giant planets with increasing separation, with measured values falling between 10$-$2\% for separations of 30$-$100 au, and 95\% confidence-level (CL) upper limits of $\lesssim$45$-$8\% for planets on 30$-$100 au orbits, and $\lesssim$5\% between 200$-$500 au. These values are in excellent agreement with recent surveys of A-stars in the solar neighborhood$-$supporting findings that giant planets out to separations of $\lesssim$100 au are more frequent around A-stars than around solar-type hosts. Finally, the relatively low occurrence rate of super-Jupiters on wide orbits, the positive correlation with stellar mass, and the inverse correlation with orbital separation are consistent with core accretion being their dominant formation mechanism.%$-$suggesting that the maximum companion mass likely scales with host star (and protoplanetary disk) mass.
\end{abstract}

%1/27 = 3.7% +/- 3.1 (68%) or +7.1/-3.6 (95%)
%$\leq$10.8\% with 95\% confidence, and between 0.6$-$6.8\% with 68\% confidence.

\section{Introduction}

%layout:

%what is the point of this paper, anyways? A-stars! Rephrase the first two paragraphs to introduce the field of direct imaging by its first few discoveries, the motivation for A-stars, and the restuls of recent surveys (including comp. to sun-like stars).
% and a significant motivation remains to identify systems with a greater likelihood of hosting directly detectable planetary systems
High-contrast imaging enables studying wide-orbit giant planets around young stars. However, the number of directly imaged planets remains low (on the order of tens). After hundreds of nearby sun-like (FGK) stars were observed over multiple surveys reporting null detections  (e.g., \citealt{Chauvin2003, Biller2007, Kasper2007, Apai2008, Nielsen2013}), the first exoplanets discovered by direct imaging$-$including the four super-Jovian planets orbiting the A5V star HR 8799 \citep{Marois2008,Marois2010} and the giant planet orbiting the A6V star $\beta$ Pictoris \citep{Lagrange2010}$-$seemed to indicate that giant planets on orbits of $\sim$10 au or greater are more common around higher-mass stars. 

If the cores of wide-separation giant planets are typically assembled via the slow, step-wise process of core accretion (e.g., \citealt{Pollack1996}), which strongly depends on the local surface density of solids and orbital timescales in protoplanetary disks (and hence on stellar mass: e.g., \citealt{Pascucci2016}), then wide-orbit giant planets should be more frequent around higher mass stars. This is indeed true for close-in giant planets (e.g., \citealt{Ghezzi2018}), and this trend should extend to larger radii if the majority of wide-orbit giant planets form similarly via core accretion. The systems mentioned above were followed up by discoveries of several other planets around A-type stars$-$including those orbiting HD 95086 \citep{Rameau2013}, HD 106906b \citep{Bailey2014}, 51 Eri \citep{Macintosh2015}, and HIP65426 \citep{Chauvin2017}. It's also worth noting that an increasing number of planets have been imaged around lower-mass stars$-$including the two planets orbiting the K7 (0.76 M$_{\odot}$) star PDS 70 \citep{Keppler2018, Wagner2018b, Haffert2019} and three planets around K-stars from the Young Suns Exoplanet Survey (YSES: \citealt{Bohn2020,Bohn2021}). While the number of imaged planets around FGK stars has increased in recent years, there remain fewer than those around A-type stars. Considering even lower-mass (M-dwarf) stellar hosts, \cite{Lannier2016} also find a lower occurrence rate of wide-orbit giant planets compared to those around A-stars.

The recent Gemini Planet Imager Exoplanet Survey (GPIES: \citealt{Nielsen2019}) presented interim results that indicate a significant ($>$99\% confidence level, or CL) difference between the wide-orbit giant planet populations of high-mass and low-mass stars. For planets between 5$-$13 M$_{Jup}$ and between $a$=10$-$100 au orbiting $>$1.5 M$_{\odot}$ stars, they found a 68\%-CL interval of 5.3$-$13.9\%; whereas similar planets orbiting stars of any spectral type yielded a 68\%-CL interval of 2.1$-$5.4\%. Similarly, the SHINE survey on VLT/SPHERE \citep{Vigan2021} that includes many of the same targets yielded a 68\%-CL interval of 13.3$-$36.5\% for the frequency of 1$-$75 M$_{Jup}$ and between $a$=5$-$300 au orbiting B/A-type stars\footnote{Note that the larger value is reflective of the larger range of parameters, particularly at lower masses and semi-major axes for which planets are more frequent.} and 3.0$–$10.5\% for planets of similar properties orbiting FGK stars. Both samples included $>$50 BA-type stars ($>$70 for GPIES) among the nearby moving groups. 

 The meta-analysis of directly imaged planets and brown dwarf companions by \cite{Wagner2019} including all known systems with detected companions also found a steeper relative companion mass function for high-mass stellar hosts compared to low-mass stellar hosts. However, the trend may not extend to all separations. Considering companions on wider orbits of 30$-$300 au, \cite{Bowler2016} found no significant difference between giant planet populations around high-mass stars and low-mass stars ($2.8_{-2.3}^{+3.7}$ \% of BA stars, $<$4.1\% of FGK stars, and $<$3.9\% of M dwarfs). Nevertheless, at separations of $\lesssim$100 au, the message appears clear: higher-mass stars have a higher frequency of hosting giant planets. 
 
The higher frequency of wide-orbit giant planets around A-stars influences models of planet formation and the target lists (and expected yields) of future surveys. However, the occurrence rate of wide-orbit giant planets around high-mass stars remains significantly uncertain. Upper-limits range from $\lesssim$20\% (GPIES: 95\%-CL, 5$-$13 M$_{Jup}$, 10$-$100 AU, \citealt{Nielsen2019}) to $\lesssim$36.5\% (SHINE: 68\%-CL, 1$-$75 M$_{Jup}$, 5$-$300 au, \citealt{Vigan2021}). These state-of-the-art surveys utilize the latest in extreme adaptive optics (ExAO) technology. However, they were also designed (with good reason) to maximize discoveries of planets on shorter periods by focusing on nearby stars. This limits the available number of A-stars, and particularly those with well-constrained ages. Well-constrained ages translate to better constraints on planetary masses (and upper limits) because planetary luminosity is correlated with both mass and age.

An opportunity to better constrain the demographics of wide-orbit giant planets around A-type stars has recently arisen with the advent of ExAO systems on large ground-based telescopes (e.g., \citealt{Macintosh2014, Jovanovic2015, Males2018, Beuzit2019}). These systems routinely reach contrasts of $\sim$10$^{-6}$ for stars within young moving groups at distances of $\gtrsim$100 pc, which enables homogeneous surveys of a large number of A-stars with well-constrained ages. The Sco OB2 association (also known as Scorpius-Centaurus-Lupus, Sco-Cen, or SCL) at $\sim$100$-$200 pc contains the largest number of nearby age-dated A-stars \citep{deZeeuw1999,Kouwenhoven2005,Kouwenhoven2007}. The ages of the stars within Sco OB2 vary between $\sim$5$-$17 Myr, depending on position within the cluster \citep{Pecaut2016}. At such young ages, the population of recently formed giant planets is luminous enough to be imaged down to $\sim$4$-$5 M$_{Jup}$ at separations $\gtrsim$30 au (e.g., \citealt{Fortney2008,Baraffe2015, Marleau2019b}).

The goal of this survey is to directly constrain the demographics of wide-orbit giant planets around the A-stars of the Sco OB2 association in order to inform planet formation models (e.g., \citealt{Forgan2018, Emsenhuber2020}), the target selections and expected yields of future surveys, and other areas where this fundamental parameter is relevant. We also aim to discover additional planets and substellar companions whose orbits and atmospheres are accessible to direct characterization. Finally, the results of this survey will be directly comparable to the demographics of planets around A-stars in the closer (but slightly older) moving groups targeted largely by past surveys (e.g., \citealt{Nielsen2019,Vigan2021}), and also to lower-mass members of the Sco OB2 association \citep{Bohn2020,Bohn2021}$-$enabling trends in exoplanet properties to be identified across host star age, mass, and formation environment.

\section{Observations and Data Reduction}

Our observations utilized the Spectro-Polarimetric High-Contrast Exoplanet Research Experiment (SPHERE) instrument on the Very Large Telescope (VLT) in Chile (\citealt{Beuzit2019}). SPHERE provides contrasts of 10$^{-5}-10^{-6}$ at separations of $\sim$1" in under an hour for $\sim$7$-$8 mag stars. SPHERE also provides diffraction-limited imaging with a resolution of $\sim$0$\farcs$05 across wavelengths of $\sim$1$-$2.2 $\mu$m. Dual-band imaging in a selection of narrow-band filters over a field of view of $\sim$6" in radius (IRDIS: \citealt{Vigan2010}) is combined with simultaneous integral field spectroscopy from 0.95$-$1.65 $\mu$m over a smaller field of view of 0$\farcs$8 in radius (IFS: \citealt{Claudi2008}). The latter enables spectral identification of close-in exoplanet candidates, while the dual-band imaging enables identifying companions on wider orbits through their common proper motion. 

Our observations typically utilized the IRDIFS\_Ext mode of SPHERE, in which dual-band images are acquired with IRDIS in the $K12$ filters ($\lambda_{K1}$=2.110$\pm$0.051 $\mu$m, $\lambda_{K2}$=2.251$\pm$0.055 $\mu$m) and spatially-resolved spectra from $Y-H$ band (0.95$-$1.65 $\mu$m, $R\sim$30) are acquired with the IFS. For some of the targets (primarily among the systems that were observed by SPHERE for other programs) the IRDIFS mode was used. This includes the $H23$ dual-band filter combination ($\lambda_{H2}$=1.593$\pm$0.026 $\mu$m, $\lambda_{H3}$=1.667$\pm$0.027 $\mu$m) and IFS spectroscopy from $Y-J$ (0.95$-$1.35 $\mu$m, $R\sim$50). Each target was observed for approximately $\sim$30$-$60 minutes of exposure time (see Appendix A), enabling field rotation of $\sim$20$-$30$^{\circ}$, and resulting in similar performance for each target.

%\startlongtable
\begin{deluxetable*}{cccccccccccc}
\tablecaption{Substellar Companions Imaged around Sco-Cen A-type Stars in This Survey}%\tablenotemark{a} \label{tab:table}}
\tablehead{
\colhead{Name} & \colhead{Sub-} & \colhead{Age} & \colhead{Proj.} & \colhead{Mass} & \colhead{$a$} & \colhead{$e$} & \colhead{$i$} & \colhead{Dist.$^{\mathrm{a}}$}& \colhead{Host} & \colhead{SpT} & \colhead{References}\\
\colhead{} & \colhead{group} & \colhead{(Myr)} & \colhead{Sep.} & \colhead{(M$_{Jup}$)}& \colhead{}& \colhead{}& \colhead{} & \colhead{(pc)}& \colhead{SpT}} & \colhead{}
%\colnumbers
\startdata
HIP75056Ab & UCL & $\sim$12 & 0$\farcs$15 & 25$\pm$5 & 30$\pm$15 au & 0.5$\pm$0.2 & 23$\pm$11$^\circ$ & 126$\pm$2 & A2V & M6$-$L2 & b \\
HD 95086b & LCC & 17$\pm$4 & 0$\farcs$6 & 4$-$5 & 59$^{+10}_{-13}$ & 0.14$^{+0.07}_{-0.14}$& 150$^{+12}_{-13}{}^{\circ}$ & 86.4$\pm$0.4 & A8V & L1$-$T3 & c, d, e \\
HIP65426b & LCC & 14$\pm$4 &  0$\farcs$8 & 6$-$12  & 94$^{+28}_{-45}$ &0.55$^{+0.42}_{-0.22}$& 112$^{+14}_{-18}{}^{\circ}$ &109.2$\pm$0.8 & A2V & L5$-$L7 & ~e, f
\label{tab:1} 
\enddata
\tablenotetext{a}{\cite{GaiaDR2}, $^{\mathrm{b}}$ \cite{Wagner2020}, $^{\mathrm{c}}$ \cite{Rameau2013}, $^{\mathrm{d}}$ \cite{DeRosa2016}, $^{\mathrm{e}}$ \cite{Bowler2020}, $^{\mathrm{f}}$ \cite{Chauvin2017}.}

%\tablenotetext{a}{\cite{Chauvin2017}}
%\tablecomments{...}
\end{deluxetable*}

\begin{figure*}[htpb]
%\figurenum{2}
\epsscale{1.15}
\plotone{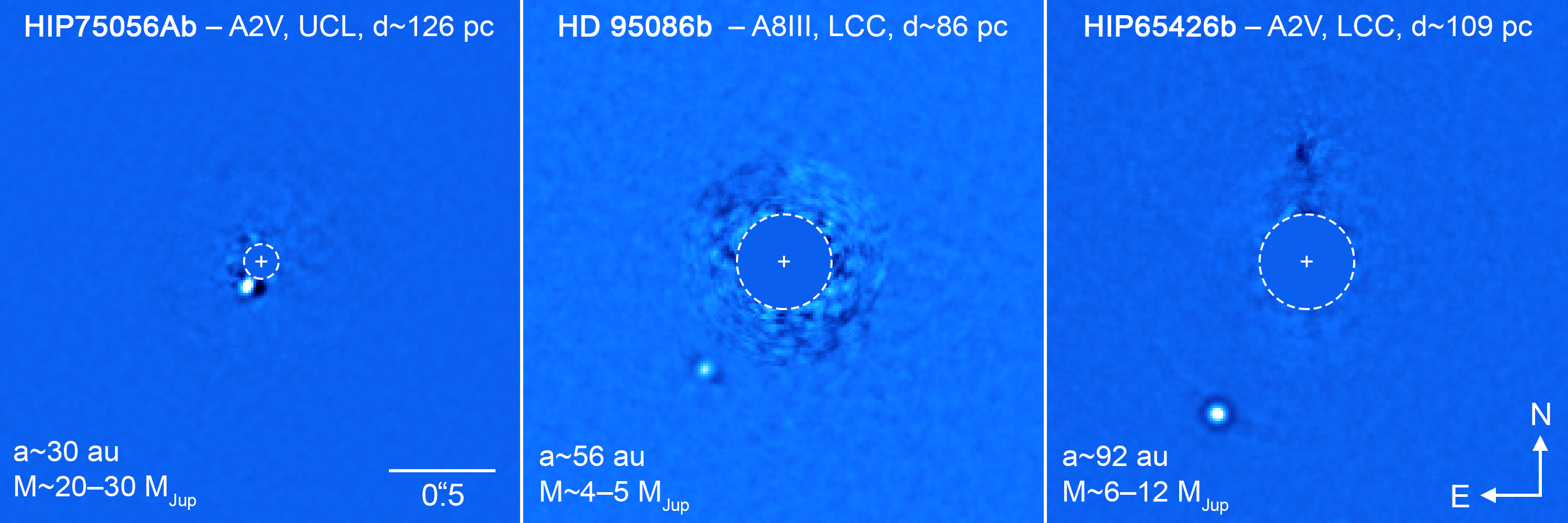}
\caption{\label{fig:1}Confirmed substellar companions among Sco-Cen A-type stars. These three objects span masses of $\sim$4$-$30 M$_{Jup}$ and separations of $\sim$15$-$100 au. The discoveries and initial analyses of HIP75056Ab, HD 95086b, and HIP65426b can be found in \cite{Wagner2020}, \cite{Rameau2013}, and \cite{Chauvin2017}, respectively.}
\end{figure*}

\subsection{Target Selection}

Targets were selected from the A-stars within the Sco OB2 association, which contains the largest group of nearby A-stars with well-determined ages \citep{deZeeuw1999}. Typical ages and distances to member systems range from $\sim$10$-$17 Myr and $\sim$100$-$200 pc \citep{Pecaut2016, GaiaDR2}. The list of A-stars among this association is thought to be complete. However, A-stars have a high ($\sim$50\%) binary fraction, which limits the available orbital phase space. \cite{Kouwenhoven2005} and \cite{Kouwenhoven2007} performed an early AO survey of these systems using the ADONIS system on the 3.6m New Technology Telescope. This survey established which A-stars host stellar companions at projected separations of $\gtrsim$0$\farcs$2. 

From the list of 115 A-star systems in \cite{Kouwenhoven2005,Kouwenhoven2007}, we selected those that may be capable of hosting wide-orbit planetary systems. The gravitational influence of a companion star will restrict the range of stable planetary orbits to those that are $\lesssim$25$-$30\% of the stellar companion's periapsis \citep{Holman1999}. For distances of $\sim$100 pc, a planet on a stable $\sim$50 au orbit translates to a minimum binary separation of $\gtrsim$1$\farcs$5. We removed those with smaller projected separations from the list, which resulted in 97 systems. Among these, we observed 63 throughout 2015$-$2019, and obtained publicly available datasets from the European Southern Observatory (ESO) archive for another 25 systems. This resulted in a total survey size of 88 Sco-Cen A-stars. Of these observations, four were deemed of inferior quality, typically due to very low field rotation ($\lesssim$5$^\circ$), and were excluded from the analysis. These include targets with Survey-ID S26, S57, S58, and SA29.

\subsection{Automated Data Reduction Pipeline}

We reduced the data for each target in a uniform manner with a self-developed automated pipeline designed to take raw data products from the ESO archive and to assemble the high-level scientific products.\footnote{Our open source pipeline is publicly available at \url{https://github.com/astrowagner/sphere-tools/releases/tag/v1.2}.} For single stars, this pipeline works well with no intervention. However, some intervention is needed in the case of wide binary systems (i.e., to mask the companion) or observations taken with a non-standard approach (e.g., those lacking standard calibrations). The pipeline results in point-spread function (PSF) subtracted images, signal to noise ratio (SNR) maps generated according to \cite{Mawet2014}, contrast curves generated through synthetic planet injections (with a SNR threshold set to 5), and lists of candidate sources above a given SNR threshold (set to 3.5 for initial identification and vetting). 

The data reduction pipeline began with the procedures presented in \cite{Apai2016} and was improved sequentially throughout the survey \citep{Wagner2015,Wagner2016,Wagner2018,Wagner2020}. The pipeline is briefly summarized here. Basic calibrations, including dark subtraction, flat field division, and determination of the coronagraphic centering via satellite spots (typically taken at the start of the observation), were performed for all datasets. For relevant epochs (those prior to 2016 July), the time synchronization correction of \cite{Maire2016} was applied to the parallactic angle information, and for all epochs the \cite{Maire2016} astrometric and field distortion corrections were applied. For single-star targets, the images were then aligned in the pupil-stabilized orientation via cross-correlation. For multi-star systems, or those with bright background contaminants, the cross-correlation was either performed on only a central image patch (for widely separated companions) or in the field-stabilized orientation for more closely separated binaries in which the wings of the PSF interfere with alignment. 

At this stage (when relevant) synthetic point sources were injected into the data using unsaturated frames taken through a neutral density filter with the target slewed $\sim$0$\farcs$5 off of the coronagraph. The PSF was then subtracted via two independent means: via classical angular differential imaging (cADI: \citealt{Marois2006}) and via projection onto eigen-images via the Karhunen-Lo\`eve Image Processing (KLIP) algorithm \citep{Soummer2012}. Given the difference in sensitivity to typical exoplanet spectral features between the K1 and K2 (H2 and H3) images, the K1 (H2) images were utilized for the subsequent analysis, except where color-based information is relevant (i.e., vetting of initial candidates).

\begin{figure*}[htpb]
%\figurenum{2}
\epsscale{1.16}
\plottwo{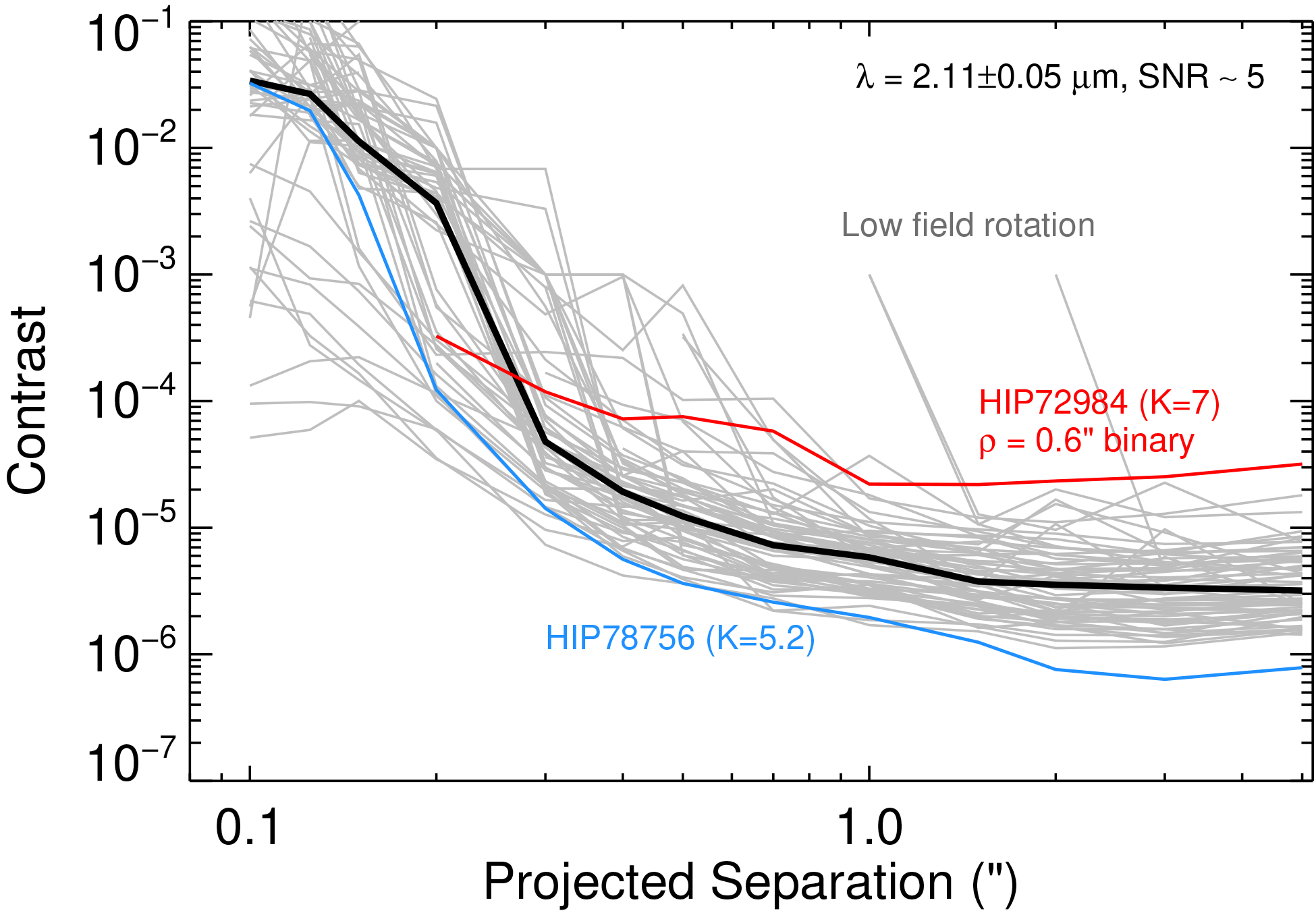}{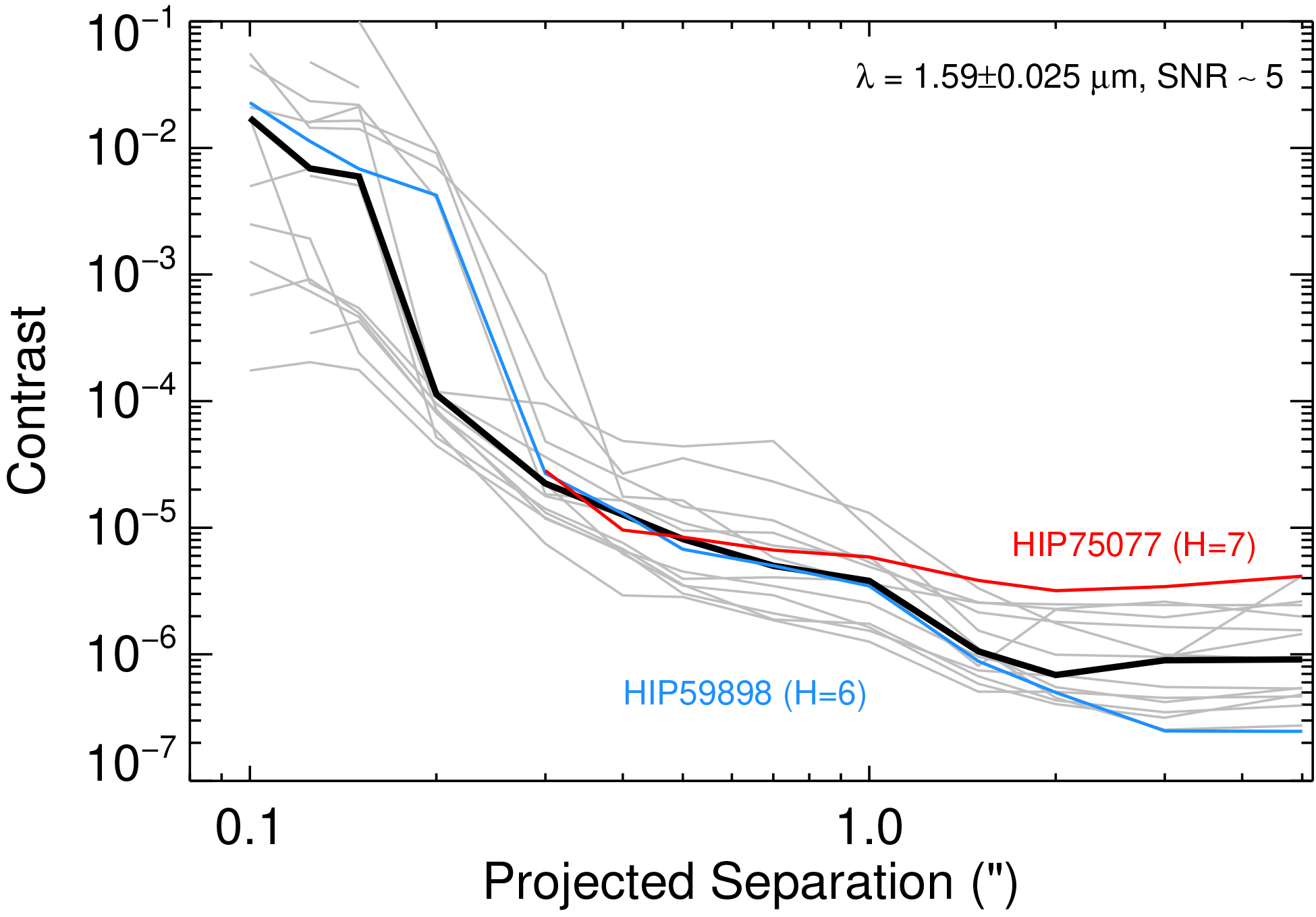}
\caption{\label{fig:2}Left: Sensitivity of the Scorpion Planet Survey in K1-band for IRDIFS\_Ext observations. Right: Sensitivity in H2-band for IRDIFS observations. In both panels the thick black curve shows the median sensitivity, while the blue and red curves show the best and worst cases for background-limited sensitivity.}
\end{figure*}

\subsubsection{Candidate Identification and Follow-up Strategy}

Candidates were identified both visually and via an automated SNR-based approach. For each pixel, we measure the flux in an aperture of diameter equivalent to the full width at half maximum (FWHM), and noise in non-overlapping apertures at the same radius. We exclude immediately adjacent apertures to minimize the impact of negative side-lobes of true signals introduced by the ADI-based PSF subtraction. We then utilize Equation 9 of \cite{Mawet2014} to calculate the SNR. Systems with candidates of SNR$\geq$3.5 were included in subsequent follow-up imaging. Many systems are near the galactic plane and hence the background contamination fraction is relatively high ($\gtrsim$50\% of the targets have at least one background object within the field of view). Ultimately, we attempted to follow-up all observations with plausible exoplanet candidates and were able to re-observe 35 out of 40 systems with identified candidates. One candidate was verified as co-moving (HIP75056Ab, \citealt{Wagner2020}), and two were identified as known planetary companions (HD 96085b and HIP65426b, \citealt{Rameau2013} and \citealt{Chauvin2017}). These are shown in Figure \ref{fig:1} and described in Table \ref{tab:1}. The rest were identified as background objects by their relative proper motion. 

For the five systems (S54, SA6, SA10, SA11, and SA25) with detected sources that were not re-observed (mostly within crowded fields and at separations greater than several arcsec), we consider these sources as likely background stars and adjusted the detection limits according to the brightest ambiguous source. A more sophisticated treatment of the systems that were not followed up would incorporate additional parameters for the possibility that the detected sources may be bound. However, given the nature of these sources and the fact that they lie in a region of parameter space that other studies have probed with greater sensitivity, these prior probabilities are close to zero. We also confirmed that removing these systems completely does not significantly alter the results.

\subsection{Sensitivity Analysis}

 We assessed the sensitivity of the data through simulated companion injection and retrieval tests. After image alignment (but prior to high-pass filtering and PSF subtraction) we injected scaled versions of the mean of the off-axis flux calibration frames\footnote{Taken with a neutral density filter (typically ND1.0 or ND2.0, which transmit $\sim$10\% and $\sim$1\%, respectively). See \url{https://www.eso.org/sci/facilities/paranal/instruments/sphere/inst/filters.html} for more details.} and proceeded with the following steps of the data reduction. We blocked regions contaminated by bright binary companions or background stars with NaN values, and measured the SNR at the position of injection according to \cite{Mawet2014}. We iterated upon the brightness of the source until it was recovered within 5\% of SNR=5. We repeated the process at separations from 0$\farcs$1$-$5" and at 10 evenly distributed azimuthal angles (beginning with PA=0$^\circ$), and then took the median of the sensitivity over azimuth. The results are shown in Figure \ref{fig:2}.
 
 For $K1$ datasets, we reach a median background-limited (5-$\sigma$) contrast of $\sim$3$\times$10$^{-6}$ at separations of $\gtrsim$1$\farcs$5. For $H2$ datasets, we reach a similar median background-limited contrast of $\sim$10$^{-6}$. These detection limits correspond to masses of $\sim$4 and $\sim$2 M$_{Jup}$ \citep{Baraffe2003} for an age of 15 Myr and distance of 125 pc around a $K$=6 ($H$=6.5) star.\footnote{The poorer background-limited sensitivity for $K1$ datasets is due to the higher sky background at longer wavelengths.} In the best cases, we reach contrasts $\sim$3$\times$ lower than the median background limits, enabling planets as low as $\sim$1 M$_{Jup}$ to be identified around a handful of targets. In the worst cases, the SNR=5 background limits are $\sim$2$-$3$\times$10$^{-5}$ for some of the $K1$ datasets, which corresponds to masses of $\sim$6 M$_{Jup}$. At $\sim$0$\farcs$5, we reach median-contrasts of $\sim$10$^{-5}$ in both $K1$ and $H2$, which corresponds to $\sim$5 $M_{Jup}$. Notably, we find a relatively uniform sensitivity to $\gtrsim$5 M$_{Jup}$ planets at projected separations of $\gtrsim$50 au.
 
 \subsection{Monte Carlo Modelling}

Previous studies have focused on exploring the scaling relations of planet frequency with physical parameters such as semi-major axis and mass through Markov Chain Monte Carlo (MCMC) simulations. Typically, power-laws in mass and semi-major axis distributions (and sometimes additional parameters, such as stellar mass: \citealt{Nielsen2019}) are explored in conjunction with frequency. Since each power-law is unknown, as well as the frequency, these three parameters are degenerate with one another. Here, we do not try to replicate this approach, and instead seek to constrain only the frequency of planets (in defined mass and semi-major axis bins) while remaining agnostic to the underlying scaling relations. For this, we utilize a simpler Monte Carlo (MC) simulation of our survey to translate the measured sensitivity to constraints on the underlying planet population (following the approach of \citealt{Kasper2007}). 

We assumed a normally distributed prior on host star distance, with distances obtained from parallaxes reported by \cite{GaiaDR2}. We assumed ages of 10 Myr, which is near the average age of the association \citep{Pecaut2016}. Since our sample was drawn with representation from the different sub-groups, the spread of ages among our sample is representative of the spread of ages among the Sco OB2 association ($\sim$5$-$17 Myr), which effectively incorporates the uncertainty in age. We assumed uniform prior mass and semi-major axis distributions to generate a large number of companions spanning the available parameter space. Note that the MC serves only to simulate a large number of planets to be compared to the detection limits, and the chosen prior distributions here do not significantly impact our results (we verified that this is the case by also testing assumed distributions of $f \propto$ $M^{-0.5}$, $M^{-1.0}$, and $M^{-1.5}$).

This approach is applicable at this step since we are only concerned with the ratio of detected vs. non-detected planets in relatively small pre-determined bins (compared to the range over which we later report the frequency). These bins are also defined in a manner such that the sensitivity across each bin is nearly uniform. We randomly generated planets on circular orbits with a uniform prior in the cosine of inclination (see the Appendix of \citealt{Brandt2014} and references therein for a discussion of the effects of non-zero eccentricity, which are of secondary importance). In this manner, 4,000 companions were simulated for each star with masses between M $=0.1-35M_{Jup}$ and $a=2-550$ au. This generated a sufficient level of smoothness within the individual maps$-$i.e., runs with greater numbers of simulated companions do not alter or improve the results.

We utilized the models of \cite{Baraffe1998,Baraffe2003,Baraffe2015} to convert mass to estimated luminosity. The choice of these high-entropy models is motivated by recent work (e.g., \citealt{Marleau2019b}) indicating that the low-entropy models (e.g., \citealt{Marley2007,Fortney2008}), in which a large portion of the accretion-generated heat escapes during formation, are an unlikely outcome of giant planet formation. We calculated the ratio of detected vs. non-detected companions over a pre-determined grid of mass and semi-major axis bins\footnote{The grid was defined as $a$=[1, 5, 10, 15, 20, 30, 40, 50, 60, 70, 80, 90, 100, 120, 150, 200, 300, 400, 500] au $\times$ $M$=[0.1, 0.5, 1, 2, 3, 4, 5, 6, 7, 8, 9, 10, 11, 12, 13, 14, 15, 20, 25, 30, 35] M$_{Jup}$. The widths of the bins were set to 10\% of the values at each grid point.} to estimate the completeness to planets around each star. These individual completeness maps were summed to create a total estimated completeness map for the survey$-$shown in the left panel of Figure \ref{fig:3}. In other words, this map shows the number of stars around which we would expect to have observed planets of a given mass and semi-major axis if they were uniformly present.

\subsection{Statistical Methods}

\begin{figure*}[htpb]
%\figurenum{2}
\epsscale{1.16}
\plottwo{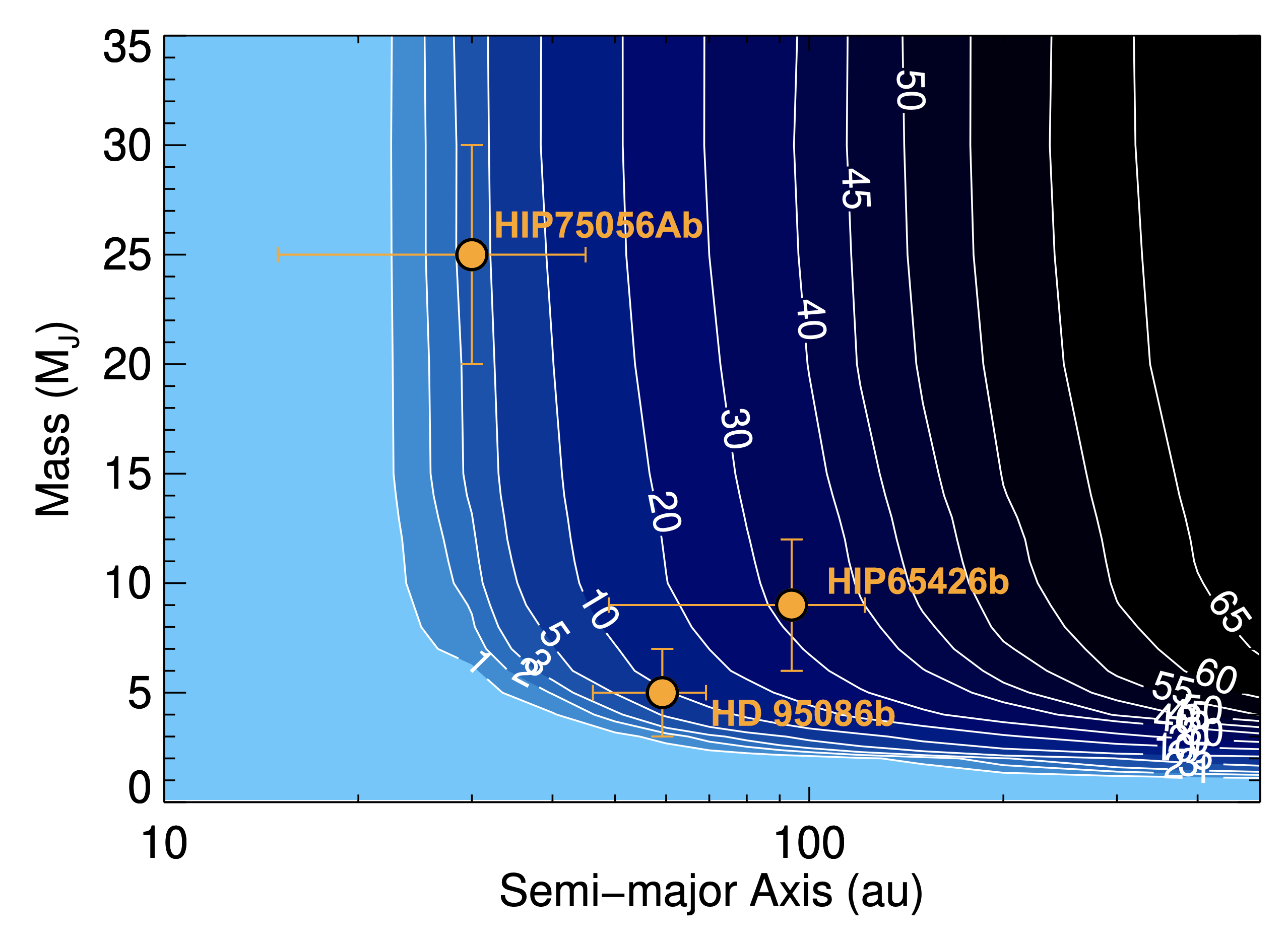}{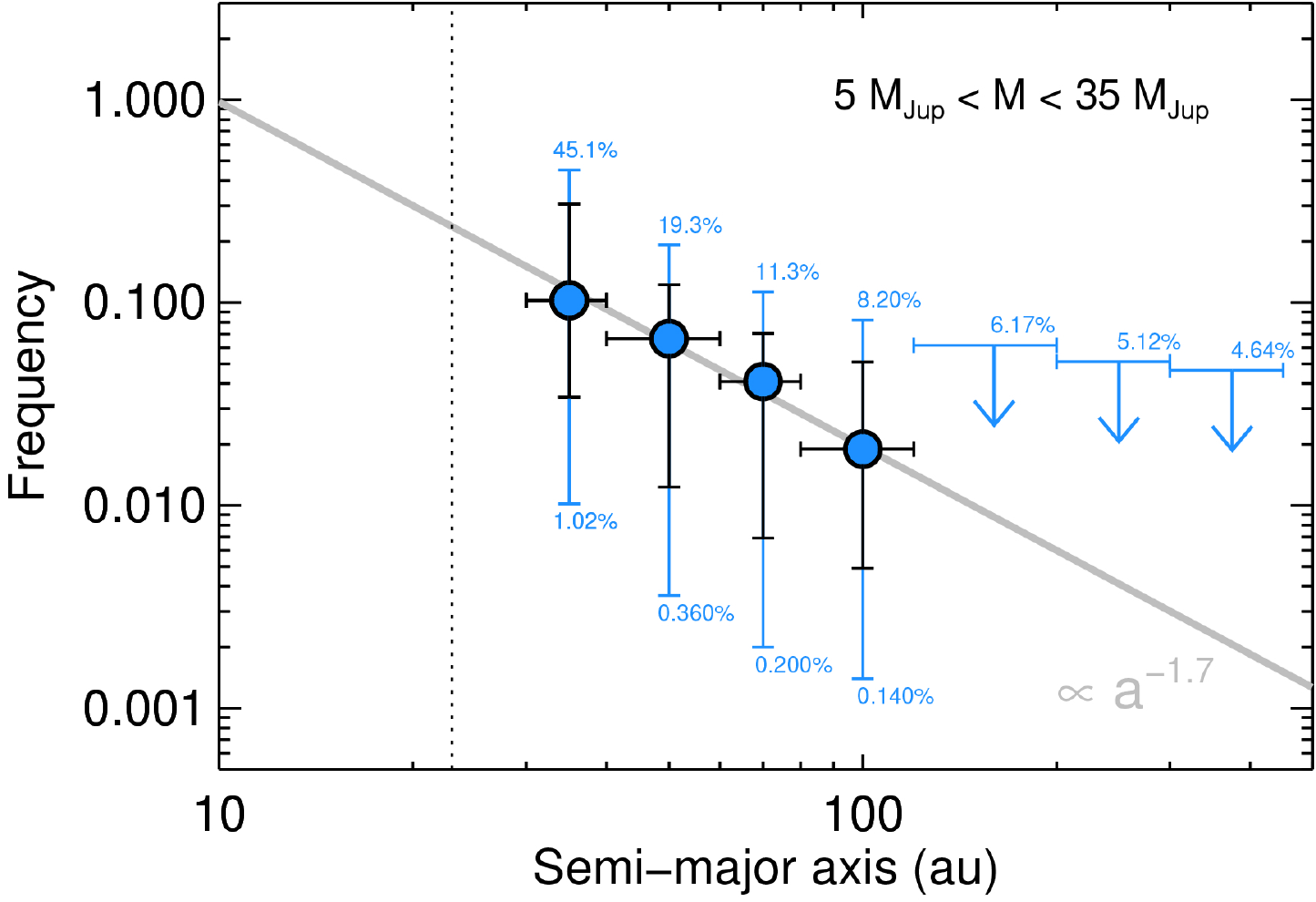}
\caption{\label{fig:3}Left: Sensitivity of the Scorpion Planet Survey from Monte Carlo simulations. The contours identify the number of stars around which a planet of a given mass and semi-major axis would have resulted in a detection if they were uniformly present. Right: Frequency vs. semi-major axis. The blue points show the measured frequency, whereas the vertical (black) uncertainty shows the 68\% confidence interval. The blue ranges denote upper and lower 95\% confidence values (i.e., their range denotes the 90\% confidence interval). The horizontal bars represent the range of the semi-major axis bins. A power-law of $f \propto a^{-1.7}$ is shown for comparison (i.e., not as a best-fit). The vertical dotted line denotes the inner-most region ($\sim$23 au) at which we would expect to have found $\geq$1 companion if they were uniformly present.}
\end{figure*}

Two-dimensional Gaussian profiles were constructed for each companion based on their available constraints on mass and semi-major axis. These were normalized such that the sum over each is equal to unity to represent a single detected companion. These  were then summed to create a detection probability map to be compared to the completeness map (Figure \ref{fig:3}, left). We consider only masses above $\sim$5 M$_{Jup}$ and separations of $a\geq$30 au in order to restrict the analysis to the region in which the sensitivity is relatively uniform. We then explored the consistency of the detections across several ranges of semi-major axes\footnote{Specifically, we assesed the frequency of giant planets of M=5$-$35 M$_{Jup}$ in semi-major axis bins of 30$-$40 au, 40$-$60 au, 60$-80$ au, 80$-$120 au, 120$-$200 au, 200$-$300 au, and 300$-$450 au. These correspond to regions with near-uniform sensitivity.} with the average expected number of stars around which we would have been capable of detecting such planets within the set range. 

We calculated confidence intervals via the binomial distribution, which gives the probability of a number of successful events  occurring among a certain number of trials with a given underlying frequency. For this, we rounded the number of detected companions and average completeness to the nearest integer. For each bin, we report the measured detection rate, along with the frequencies that are consistent with the data at both the 68\%-CL and 95\%-CL. For bins in which the number of detected companions rounds to zero (i.e., those at $>$120 au), we report only 95\%-CL upper limits (Figure \ref{fig:3}). %The results are shown in the right panel of Figure \ref{fig:3} and are discussed further in \S3.2.

\section{Results}

\subsection{Detected Companions}

While the main result of our survey is its statistical analysis, the number of detected companions is small enough to briefly consider their properties on an individual basis. The detected companions are described in Table \ref{tab:1} and shown in Figure \ref{fig:1}. They span masses of several times that of Jupiter to $\sim$25 M$_{Jup}$ and projected separations of $\sim$15$-$90 au. Their established properties and relevant references are summarized briefly below.

\subsubsection{HIP75056Ab}

HIP75056Ab is the most massive substellar companion among the sample and that of the shortest orbital period. This $\sim$20$-$30 M$_{Jup}$ companion at $\sim$0$\farcs$15 to an A2V star was discovered as part of our survey \citep{Wagner2020}. It orbits the primary component of a wide-separation (5$\farcs$2) binary with a mass ratio of $q\sim$ 0.16 within the Upper Centaurus Lupus (UCL) sub-group of Sco OB2. HIP75056Ab was detected in 2015 and 2019$-$establishing that it is co-moving with the system on an orbit of $a$=30$\pm$15 au and $e$=0.5$\pm$0.2. From IFS data combined with the companion's $K1$ and $K2$ photometry, \cite{Wagner2020} inferred a spectral type of M8$\pm$1 and a temperature of $\sim$2300$\pm$300 K. From our sensitivity and completeness analysis, we estimate that we would be sensitive to $\sim$3 such companions if they had an occurrence rate of one per star. The detection of this companion in a region of parameter space with such relatively low sensitivity is likely indicative of a relatively high frequency of companions at smaller separations, and not a higher relative frequency of more massive companions due to the sharply decreasing frequency of companions more massive than $\sim$10$-$20 M$_{Jup}$ at all separations (e.g., \citealt{Wagner2019}).

\subsubsection{HD 95086b}

HD 95086b is the least-massive companion detected in our survey. This $\sim$4$-$5 M$_{Jup}$ exoplanet was discovered by \cite{Rameau2013} with VLT/NaCo. The host is a single A8 pre-main sequence star toward the near-side of the Lower Centaurus-Crux (LCC) subgroup of the Sco OB2 association. The planet is on an orbit with $a\sim$ 40$-$70 au and $e\lesssim$ 0.2 \citep{Bowler2020}. The system also hosts a debris disk \citep{Su2017} with a background galaxy in projected separation \citep{Zapata2018}. From our sensitivity and completeness analysis, we estimate that we would have the capability to detect such a companion around $\sim$7 stars.

\subsubsection{HIP65426b}

At a projected separation of $\sim$90 au, HIP65426b is the widest-separation companion detected in our survey. This $\sim$6$-$12 M$_{Jup}$ companion was discovered by \cite{Chauvin2017} as part of the SHINE survey \citep{Vigan2021}. This companion is the most similar in properties among those detected here to the iconic HR 8799b \citep{Marois2008}, which is the type of companion whose frequency our survey was designed to constrain ($M\sim$ 5 M$_{Jup}$, $a\sim$ 70 au). From our sensitivity analysis, we estimate that we would have the capability to detect a companion such as HIP65426b around $\sim$27 stars. A trend that small-separation companions (between $\sim$5$-$30 M$_{Jup}$) are more frequent may already be apparent.

%\begin{figure*}[htpb]
%\figurenum{2}
%\epsscale{1.2}
%\plotone{detections_individual_v1.png}
%\caption{Left: }
%\end{figure*}

%\begin{figure*}[htpb]
%\figurenum{2}
%\epsscale{1.15}
%\plottwo{sma.png}{mass_W19.png}
%\caption{Left: The underlying occurrence rate (frequency) inferred from the sensitivity levels to each detected companion vs. orbital semi-major axis. Uncertainties in semi-major axis are obtained from orbital fits, whereas 1-$\sigma$ uncertainties in occurrence rate are given by the binomial distribution. A power-law fit suggests a dependence of frequency $\propto$ a$^{-1.8\pm0.8}$. Right: underlying occurrence rate vs. companion mass. A power-law of $\propto$ M$^{-1.3\pm0.3}$ (\citealt{Wagner2019}) is overlayed.}
%\end{figure*}

\subsection{Wide-Orbit Giant Planet Frequency}

This survey was designed primarily to constrain the frequency of wide-orbit super-Jupiters, with HR 8799-b at $\sim$70 au as the prime example \citep{Marois2008,Marois2010}. In other words, we are primarily focused on those between $M$=5$-$15 M$_{Jup}$ and $a=60-80$ au. Within this range, we measured a frequency of 3.4$^{+5.4}_{-2.5}$\% and establish a 95\%-CL upper limit of $\leq$13.9\%. This is somewhat lower than (but consistent with) values from other recent surveys, for reasons that will be discussed in \S4. We also explored frequency constraints at different separations and across a broader range of masses (5$-$35 M$_{Jup}$). The measurements of giant planet frequency vs. semi-major axis are shown in the right panel of Figure \ref{fig:3}. For companions with M=5$-$35 M$_{Jup}$ and $a\sim$ 30$-$40 au (80$-$120 au), we measured observed frequencies of $\sim$10\% ($\sim$2\%), with a 95\% upper-limit of $\leq$45\% ($\leq$8\%). Consistent with the results of most other surveys, we measure a decreasing frequency of companions with semi-major axis.% that is similar to a powerlaw of $N\propto a^{-1.7}$. However, this carries a significant uncertainty, and we refer to \cite{Nielsen2019} and \cite{Vigan2021} for a proper treatment of constraints and uncertainties on this relation. 

\subsection{Protoplanetary and Debris Disks}

While our observations were designed to directly image exoplanets, they are also sensitive to scattered light from circumstellar dust. Here, we briefly and qualitatively describe those results. We obtained the first images of the debris disk around HD 110058 \citep{Kasper2015} and the spiral protoplanetary disk around HD 100453 \citep{Wagner2015}, which were both inferred to exist through their infrared spectral energy distributions (SEDs).\footnote{However, we did not detect several other disks among our sample that were also identified through their SEDs.} The presence of such disks (especially the relatively massive disk round HD 100453) is evidence of the youth of Sco OB2, which is in line with recent estimates of the maximum age of $\lesssim$17 Myr \citep{Pecaut2016}. While these disks lower the sensitivity to planets within these systems (accounted for via synthetic injections), they also open additional opportunities for planet formation research (e.g., \citealt{Dong2016,Wagner2018, Nealon2020}), through, for instance, studying disk structures as tracers of dynamical processes.
%Maybe this should be an appendix? 

\subsection{HD 131399Ab}

\begin{figure}[htpb]
%\figurenum{2}
\epsscale{1.15}
\plotone{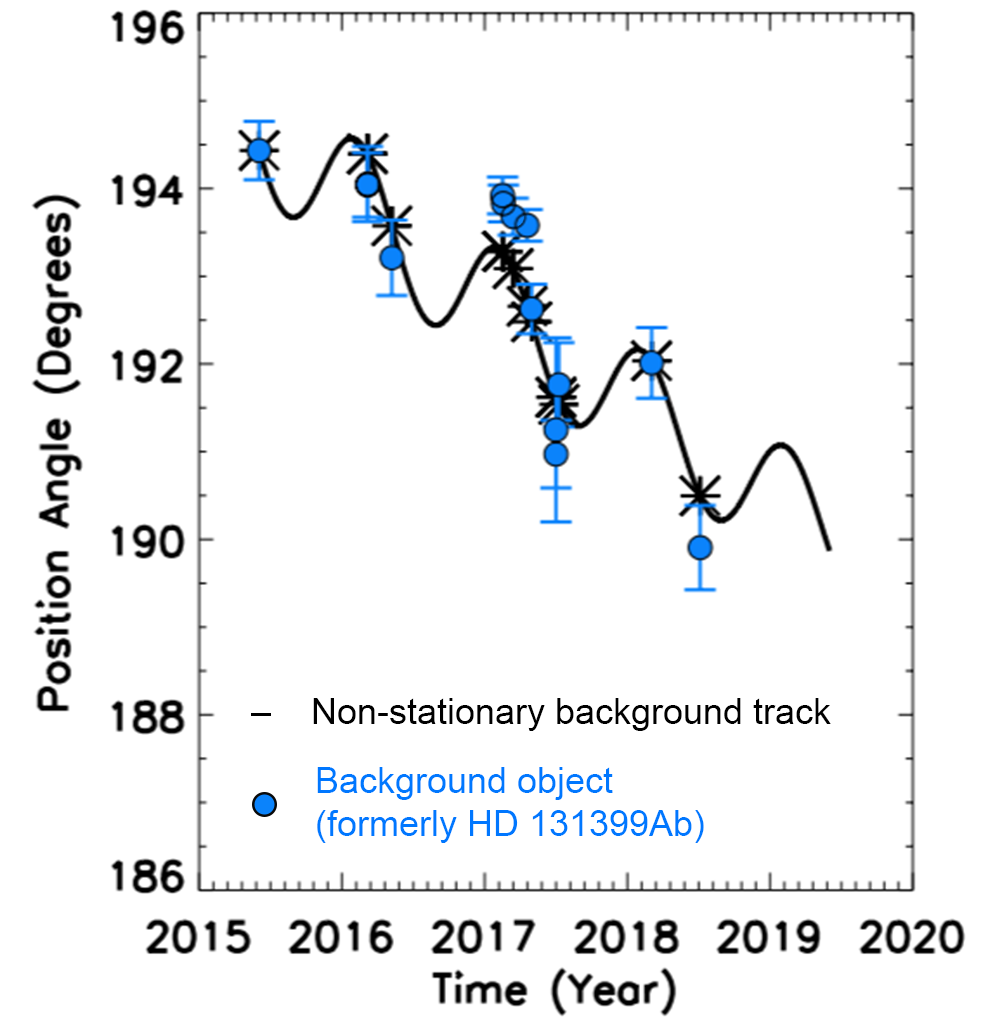}
\caption{\label{fig:4}Evolution of relative position angle vs. time of the background object near HD 131399A. The response to the foreground system's (HD 131399's) greater parallax is most notably apparent in the 2017 and 2018 years, which confirms this previous exoplanet candidate as a background contaminant.}
\end{figure}

In \cite{Wagner2016}, we reported the detection of a planetary-mass companion on an $\sim$80 au orbit within the HD 131399 quadruple system (at the time thought to be a triple: \citealt{Lagrange2017}). Subsequent observations determined that this object is likely to be a high-velocity background star with its own significant proper motion aligned with the system \citep{Nielsen2017}. Here, we report additional observations in support of this hypothesis. Our extended time baseline covers several years and displays a clear parallax difference between this object and the HD 131399 system that is most notably seen in the plot of position angle vs. time (Figure \ref{fig:4}). This confirms that it is indeed a background object. HD 131399 and its planet-hosting classification history carry a useful lesson, as assumptions of a stationary background are still commonly assumed. The background object has an approximate proper motion of $-$5 mas/yr and $-$11 mas/yr in right ascension and declination, respectively, and an parallax of $\lesssim$1 mas (i.e., a distance of $\gtrsim$1 kpc and a 2D relative velocity of $\gtrsim$57 km/s with respect to the Sun). This motion is aligned with the system such that initial common proper motion tests yielded an incorrect classification of the object as a likely bound companion.

\section{Discussion}

The primary scientific motivation of this survey is to constrain the frequency of wide-orbit super-Jovian planets orbiting A-stars. Few planet formation models, if any, predict a significant ($\gtrsim$10\%) occurrence rate of $\sim$5$-$15 M$_{Jup}$ planets at $\sim$70 au around hosts of any stellar type. Therefore, this parameter is a fundamental test of the existing models (e.g., \citealt{Pollack1996,Boss1997,Mordasini2009,Mordasini2012,Forgan2018,Emsenhuber2020}, etc.) and one that is becoming better constrained as the instrumentation progresses. 

The frequency of wide-orbit giant planets around A-type stars is also relevant to determine the future yield of exoplanet imaging surveys. For this, trends with separation play a crucial role, as more parameter space is becoming accessible at smaller separations as AO technology improves along with increasing primary mirror diameters. The evidence so far suggests quite strongly that giant planet occurrence rates increase toward smaller separations (until a few au: \citealt{Fernandes2019,Fulton2021}). However, the frequency of such companions remains largely uncertain. This is often the reason why surveys report occurrence rates over a wide range of parameters. In order to utilize a survey's best sensitivity, constraints must necessarily come from the outer regions ($\gtrsim$30 au). Often, ranges for reported frequencies extend over hundreds of au, and are easily misinterpreted as implying greater precision than actually exists for planets at the smaller ends of the semi-major axis and mass ranges.

Recently, \cite{Nielsen2019} and \cite{Vigan2021} reported frequencies for giant planets orbiting BA-stars of 9$^{+5}_{-4}$\% (M=5$-$13 M$_{Jup}$, $a$=10$-$100 au), and 23.0$^{+13.5}_{-9.7}$\% (M=1$-$75 M$_{Jup}$, $a$=5$-$300 au), respectively. Over the same range as \cite{Nielsen2019}, we measure an occurrence rate of 12.8$^{+10.4}_{-7.2}$\%. These relatively high estimates and uncertainties are reflective of the ranges involved and the higher frequency of planets with small masses and at small separations. The meta-analysis of \cite{Bowler2016}, which considered only companions on $a\geq$30 au orbits and found a relatively low occurrence rate of wide-orbit planets orbiting BA-stars ($\sim$3\%), is also indicative of the bias of higher occurrence rates at small separations. 

Considering  planets with orbital semimajor axes of $a$=60$-$80 au and masses of M$=$5$-$15 M$_{Jup}$, we find an occurrence rate of 3.4$^{+5.4}_{-2.5}$\% and a 95\%-CL upper limit of $<$13.9\%. In comparison, for M$=$5$-$35 M$_{Jup}$ companions at separations of $a=$30$-$40 au (i.e., those at shorter orbital periods and including larger masses), we measure an occurrence rate of $\sim$10\% and establish a 95\%-CL upper limit of $\lesssim$45\%. This illustrates the power of establishing sensitivity at smaller separations to dramatically increase the yield of direct imaging planet surveys. These results are also consistent with the majority of directly imaged planetary-mass companions at a$\lesssim$100 au forming via core accretion (e.g., \citealt{Kratter2010,Wagner2019}). In contrast, giant planets formed via gravitational instability should be more common at larger radii and around less massive stars, as instabilities are more likely in the cold outer disk regions (e.g., \citealt{Boss1997}).

Finally, our results offer a complementary sample to past studies that were largely focused on lower-mass stellar hosts in the solar neighborhood, and also to other on-going surveys of both higher and lower-mass stars in Sco OB2. Taken in combination with GPIES \citep{Nielsen2019} and SHINE \citep{Vigan2021}, the BEASTS survey of B-type stars \citep{Janson2021}, and YSES survey \citep{Bohn2020,Bohn2021} of solar-type stars (both also focused on Sco OB2), these surveys provide a rich and comprehensive view of wide-orbit giant plants around nearby young stars. Together, they inform our understanding of exoplanet demographics, and perhaps of equal importance, enable the yield of future exoplanet imaging surveys to be enhanced by targeting stars around which detectable planets are more likely to exist.

\section{Summary and Conclusions}

We conducted a uniform blind survey for substellar companoions around 88 A-type stars in the Scorpius-Centaurus OB association with the VLT/SPHERE extreme-AO system. We detected three substellar companions$-$one of which, HIP75056Ab, marks a new discovery. We computed the sensitivity to companions across a two-dimensional grid of separations and position angles with forward modelling of injected (simulated) planetary signals in the data, and utilized these results to explore the consistency with underlying planet populations through Monte Carlo simulations. 

We find that giant planets on wide orbits (like HR 8799b and HIP65426b) with masses of 5$-$15 M$_{Jup}$ and semi-major axes of 60$-$80 au have an occurrence rate of 3.4$^{+5.4}_{-2.5}$\% and a 95\%-CL upper limit of $\leq$13.9\%. For companions between M=5$-$35 M$_{Jup}$, we measure frequencies between 10$-$2\% for separations of 30$-$100 au, and 95\% confidence-level (CL) upper limits of $\lesssim$45$-$8\% for planets on 30$-$100 au orbits, and $\lesssim$5\% between 200$-$500 au. In line with other recent studies (e.g., \citealt{Nielsen2019,Wagner2019,Vigan2021}), we confirm the trend of increasing occurrence rates as a function of smaller separations. When compared to other studies of less-massive stars (e.g., \citealt{Chauvin2003,Biller2007,Kasper2007,Stone2018,Nielsen2019,Vigan2021}), we confirm a trend of more massive stars having a higher frequency of wide-orbit giant planets. This is consistent with exoplanet population trends among inner planetary systems (e.g., \citealt{Ghezzi2018, Mulders2021}) and also with core accretion being the dominant formation mechanism among this population (e.g., \citealt{Kratter2010,Wagner2019}). Consequently, further progress to enable sensitivity to high contrast ratios at small separations and a dedicated focus on young A-type stars should significantly raise the yield of exoplanet imaging surveys.

%Additionally, we present further astrometric evidence that HD 131399Ab, a previously reported exoplanet on a wide orbit within a triple system \citep{Wagner2016}, is in reality a background contaminant with a significant intrinsic proper motion \citep{Nielsen2017}.

\section{Acknowledgments} This work is based on observations performed with ESO's Very Large Telescope and SPHERE instrument under program IDs 095.C-0298, 095.C-0389, 096.C-0241, 296.C-5036, 097.C-0545, 097.C-0060, 097.C-0826, 097.C-0865, 097.C-0949, 097.C-1019, 099.C-0247, 099.C-0300, 099.C-0402, 0101.C-0513, 0103.C-0628, 198.C-0209, and 1100.C-0481. The results reported herein benefited from collaborations and/or information exchange within NASA's Nexus for Exoplanet System Science (NExSS) research coordination network sponsored by NASA's Science Mission Directorate. K.W. acknowledges support from NASA through the NASA Hubble Fellowship grant HST-HF2-51472.001-A awarded by the Space Telescope Science Institute, which is operated by the Association of Universities for Research in Astronomy, Incorporated, under NASA contract NAS5-26555.

%\appendix
\newpage
\section{Appendix A: Target List}

\begin{deluxetable*}{ccccccccccccc}
%\startlongtable

\tablecaption{Targets $-$ Sco-Cen A-type Stars}%\tablenotemark{a} \label{tab:table}}
\tablehead{
\colhead{Target} & \colhead{R.A.} & \colhead{Dec.} & \colhead{HIP \#} & \colhead{$K_{s}$} & \colhead{SpT} & \colhead{Sub-} & \colhead{Date Obs.} & \colhead{Exp. Time} & \colhead{Field} & \colhead{Filter} & \colhead{DIT}  & \colhead{Survey}\\
\colhead{\#} & \colhead{(J2000)} & \colhead{(J2000)} & \colhead{} & \colhead{} & \colhead{} & \colhead{group} & \colhead{(YYYY-MM-DD)} & \colhead{(min)} & \colhead{Rot. ($^{\circ}$)} & \colhead{} & \colhead{(sec)} & \colhead{ID}}
%\colnumbers
\startdata
%50520 & 11 & 6.23 & A1V & LCC & & & & K\\
%53524 & A1 & 6.76 & A8III & LCC & & & &\\
%54231 & A2 & 6.75 & A0V & LCC & & & & H\\
%55188 & 41 & 7.43 & A2V & LCC & & & & K\\
%55899 & A3 & 7.07 & A0V & LCC   & & & & &  \\
%56354 & 6 & 5.78 & A9V & LCC & & & & K\\
%56993 & 44 & 7.38 & A0V & LCC & & & & K\\
%57809 & 47 & 6.61 & A0V & LCC & & & & K\\
%58465 & A5 & 6.32 & A2V & LCC & & & & H\\ 
%59282 & A6 & 7.00 & A3V & LCC   & & & & &  \\
%59397 & 51 & 7.01 & A2V & LCC & & & & K\\
1&10 19 05.1 & -64 40 35& 50520&  6.23& A1V       &         LCC  & 2015-04-26 & 42.1 & 19.1 & $K12$ & 8.00 & S11\\
2&10 57 03.0 &-68 40 02& 53524 & 6.76 &A8III      &         LCC  & 2019-04-13 & 101 & 33.8 & $K12$ & 96.0 & SA1\_2\\
3&11 05 45.7 &-47 26 32& 54231 & 6.75 &A0V        &         LCC  & 2015-02-06 & 68.3 & 39.2 & $H23$ & 64.0 & SA2\\
4&11 17 58.1 &-64 02 33& 55188 & 7.43 &A2V        &         LCC  & 2016-05-20 & 25.6 & 9.77 & $K12$ & 16.0 & S41\\
5&11 27 29.4 &-39 52 35& 55899 & 7.07 &A0V        &         LCC  & 2018-02-26 & 76.8 & 61.0 & $K12$ & 96.0 & SA3\\
6&11 33 05.6 &-54 19 29& 56354 & 5.78 &A9V        &         LCC  & 2015-04-10 & 23.5 & 12.5 & $K12$ & 8.00 & S6\\
& & &   &   &        &            & 2016-01-21 & 23.5 & 7.65 & $H23$ & 8.00 & S6\_3\\
& &&  & &  &          & 2016-01-23 & 23.5 & 7.39 & $K12$ & 8.00 & S6\_2\\
%11 40 38.6 &-49 30 33& 56963 & 7.46 &A3V        &         LCC  & & & & &  \\
7&11 41 00.2 &-54 32 56& 56993 & 7.38 &A0V        &         LCC & 2016-06-05 & 22.4 & 12.1 & $K12$ & 16.0 & S44\\
& & &   &   &        &           & 2019-04-09 & 27.9 & 15.2 & $K12$ & 8.00 & S44\_2\\
8&11 51 13.1 &-43 55 59& 57809 & 6.61 &A0V        &         LCC  & 2017-05-15 & 32.0 & 22.7 & $K12$ & 16.0 & S47\\
9&11 59 23.7 &-57 10 05 &58465 & 6.32 & A2V      &      LCC   & 2016-03-30 & 64.0 & 27.9 & $H23$ & 64.0 & SA5\\
10&12 09 38.8 &-58 20 59 &59282 & 7.00 &A3V      &      LCC   & 2017-05-04 & 68.3 & 34.8 & $K12$ & 64.0 & SA6\\
11&12 11 05.9 &-56 24 05& 59397 & 7.01& A2V      &      LCC    & 2017-06-01 & 31.5 & 14.5 & $K12$ & 16.0 & S51\\
12&12 11 14.8 &-52 13 03& 59413 & 7.46& A6V      &      LCC   & 2015-04-10 & 17.9 & 13.1 & $K12$ & 8.00 & S23\\
& & &   &   &        &           & 2015-06-16 & 25.6 & 13.5 & $K12$ & 16.0 & S23\_2\\
& &&  & &  &          & 2019-05-29 & 29.3 & 16.4 & $K12$ & 8.00 & S23\_3\\
13&12 12 10.3 &-63 27 15& 59502 & 6.80& A2V      &      LCC   & 2018-03-24 & 32.0 & 12.3 & $K12$ & 16.0 & S63\\
& & &   &   &        &            & 2019-05-26 & 29.9 & 12.3 & $K12$ & 8.00 & S63\_2\\
14&12 17 06.3 &-65 41 35& 59898 & 5.99& A0V      &      LCC    & 2016-04-06 & 67.2 & 23.9 & $H23$ & 64.0 & SA7\\
15&12 24 51.9 &-72 36 14& 60561 & 6.59& A0V      &      LCC   & 2017-06-13 & 57.6 & 24.4 & $K12$ & 96.0 & SA8\\
16&12 28 19.3 &-64 20 28& 60851 & 5.98& A0Vn     &      LCC   & 2017-05-15 & 32.0 & 12.1 & $K12$ & 16.0 & S48\\
& &&  & &  &          & 2019-04-26 & 27.5 & 12.1 & $K12$ & 8.00 & S48\_2\\
& &&  & &  &          & 2019-05-29 & 27.7 & 12.3 & $K12$ & 8.00 & S48\_3\\
17&12 33 19.9 &-54 58 52& 61265 & 7.44& A2V      &      LCC   & 2016-04-23 & 19.2 & 10.9 & $K12$ & 16.0 & S35\\
& &&  & &  &          & 2019-04-09 & 29.1 & 15.1 & $K12$ & 8.00 & S35\_2\\
18&12 38 07.3 &-55 55 52& 61639 & 6.94& A1/A2V    &     LCC    & 2017-06-01 & 31.5 & 14.9 & $K12$ & 16.0 & S52\\
19&12 39 46.2 &-49 11 56& 61782 & 7.56& A0V        &    LCC    & 2015-04-04 & 19.9 & 14.6 & $K12$ & 8.00 & S1\\
& &&  & &  &          & 2015-04-13 & 58.4 & 34.4 & $H23$ & 16.0 & S1\_2*\\
%12 56 58.2 &-54 35 14& 63204 & 6.47& A0p         &   LCC   & & & & &  SA9(Excl.) \\
20&12 57 26.2 &-67 57 39& 63236 & 6.66& A2IV/V      &   LCC   & 2017-06-15 & 51.2 & 33.9 & $K12$ & 96.0 & SA10\\
21&13 05 02.0 &-64 26 30& 63839 & 6.66& A0V         &   LCC   & 2016-04-17 & 83.2 & 30.5 & $H23$ & 64.0 & SA11\\
22&13 10 58.4 &-52 34 01& 64320 & 6.22& Ap          &   LCC   & 2015-04-09 & 21.6 & 12.8 & $K12$ & 8.00 & S4\\
& & &   &   &        &            & 2017-06-22 & 31.2 & 16.3 & $K12$ & 16.0 & S4\_2\\
23&13 18 24.9 &-45 45 53& 64925 & 6.88& A0V          &  LCC   & 2016-05-20 & 25.6 & 14.3 & $K12$ & 16.0 & S42\\
24&13 18 34.9 &-51 17 09& 64933 & 6.29& A0V          &  LCC   & 2015-05-01 & 18.8 & 13.6 & $K12$ & 8.00 & S15\\
& &&  & &  &          & 2019-05-26 & 29.9 & 17.1 & $K12$ & 8.00 & S15\_2\\
25&13 20 26.8 &-49 13 25& 65089 & 7.37& A7/A8V       &  LCC  & 2017-07-01 & 31.7 & 18.4 & $K12$ & 16.0 & S55\\
& &&  & &  &          & 2018-04-08 & 27.7 & 16.0 & $K12$ & 16.0 & S55\_2\\
26&13 21 57.1 &-51 16 56& 65219 & 6.52& A3/A4 &  LCC   & 2015-07-10 & 25.6 & 13.0 & $K12$ & 16.0 & S27\\
& &&  & &  &          & 2019-05-31 & 29.6 & 17.5 & $K12$ & 8.00 & S27\_2\\
27&13 24 08.6 &-53 47 35& 65394 & 7.25& A1Vn &       LCC   & 2018-04-10 & 76.8 & 35.5 & $K12$ & 96.0 & SA12\\
28&13 24 36.1 &-51 30 16& 65426 & 6.78& A2V     &       LCC   & 2017-02-09 & 59.7 & 49.1 & $K12$ & 64.0 & SA13\\
29&13 29 36.2 &-47 52 33& 65822 & 6.68& A1V      &      LCC     & 2015-06-11 & 24.5 & 15.1 & $K12$ & 16.0 & S20\\
30&13 32 39.2 &-44 27 01& 66068 & 7.04& A1/A2V    &     LCC   & 2017-05-07 & 25.6 & 17.5 & $K12$ & 64.0 & SA14\\
%13 37 17.8 &-40 53 52& 66447 & 7.16& A3IV/V     &    UCL   & & & & &  \\
31&13 38 42.9 &-44 30 59& 66566 & 7.36& A1V &           LCC  & 2015-05-15 & 50.1 & 33.6 & $H23$ & 64.0 & SA15\\
32&13 40 37.7 &-44 19 49& 66722 & 6.32& A0V  &          UCL   & 2018-02-22 & 32.0 & 21.9 & $K12$ & 16.0 & S60\\
& &&  & &  &          & 2019-05-25 & 20.9 & 22.7 & $K12$ & 8.00 & S60\_2\\
& &&  & &  &         & 2019-06-27 & 28.5 & 22.8 & $K12$ & 8.00 & S60\_3\\
& &&  & &  &          & 2019-06-30 & 29.6 & 21.7 & $K12$ & 16.0 & S60\_4\\
%&13 42 43.7 &-43 11 08& 66908 & 6.86& A4V   &         UCL   & & & & &  \\
33&13 44 16.0 &-51 00 45& 67036 & 6.69& A0p    &        LCC  & 2015-07-17 & 25.6 & 13.3 & $K12$ & 16.0 & S30\\
34 &13 56 20.0 &-54 07 57& 68080 & 6.28& A1V     &       UCL  & 2016-06-05 & 32.3 & 17.0 & $H23$ & 16.0 & SA17\\ %first auto-generated one, correct those above if necessary
35 &14 01 45.7 &-39 26 19& 68532 & 7.03& A3IV/V   &      UCL   & 2015-04-09 & 23.3 & 23.5 & $K12$ & 8.00 & S5\\
& &&  & &  &          & 2019-05-09 & 29.6 & 27.9 & $K12$ & 8.00 & S5\_2\\
& &&  & &  &          & 2019-05-29 & 29.5 & 29.9 & $K12$ & 8.00 & S5\_3\\
36 &14 04 42.1 &-50 04 17& 68781 & 7.38& A2V       &     UCL  & 2016-03-30 & 42.7 & 27.8 & $H23$ & 64.0 & SA18\\
37 &14 06 08.2 &-44 41 21& 68867 & 7.17& A0V        &    UCL   & 2017-06-24 & 32.0 & 22.0 & $K12$ & 16.0 & S53\\
38 &14 06 58.2 &-47 35 21& 68958 & 6.72& Ap...       &   UCL   & 2015-04-15 & 23.3 & 15.7 & $K12$ & 8.00 & S7\\
& &&  & &  &          & 2018-03-15 & 27.7 & 16.8 & $K12$ & 16.0 & S7\_2\\
39&14 24 37.0 &-47 10 40& 70441 & 7.31& A1V &           UCL   & 2016-07-23 & 42.7 & 25.2 & $H23$ & 32.0 & SA19\\
40&14 27 33.6 &-46 12 49& 70697 & 7.17& A0V  &          UCL   & 2015-05-11 & 72.5 & 51.6 & $H23$ & 64.0 & SA20\\
41&14 28 51.9 &-47 59 32& 70809 & 6.54& Ap... &         UCL   & 2017-07-24 & 38.1 & 33.1 & $K12$ & 16.0 & S56\\
42&14 29 58.4 &-56 07 52& 70904 & 6.39& A6V    &        UCL   & 2015-06-09 & 25.6 & 12.0 & $K12$ & 16.0 & S18\\
& &&  & &  &          & 2017-05-15 & 32.0 & 14.7 & $K12$ & 16.0 & S18\_2\\
43&14 30 10.0 &-43 51 50& 70918 & 6.35& A0/A1V  &       UCL   & 2017-06-24 & 30.9 & 22.9 & $K12$ & 16.0 & S54
%&14 35 05.3 &-43 33 16& 71321 & 7.17& A9V        &    UCL   & & & & &  \\

\enddata
\tablecomments{Survey IDs were assigned in numerical order as observations were completed, except for those obtained from the archive (those beginning with 'SA'), which were assigned in order of increasing R.A. (Right Ascension). The other abbreviations in the column headers are Dec. (Declination), SpT (Spectral Type), and DIT (Detector Integration Time).}
%\tablenotetext{a}{\cite{Rameau2013}}
%\tablenotetext{a}{\cite{Chauvin2017}}
%\tablecomments{...}
\end{deluxetable*}

\begin{deluxetable*}{ccccccccccccc}
%\startlongtable

\tablecaption{Targets $-$ Sco-Cen A-type Stars (Continued)}%\tablenotemark{a} \label{tab:table}}
\tablehead{
\colhead{Target} & \colhead{R.A.} & \colhead{Dec.} & \colhead{HIP \#} & \colhead{$K_{s}$} & \colhead{SpT} & \colhead{Sub-} & \colhead{Date Obs.} & \colhead{Exp. Time} & \colhead{Field} & \colhead{Filter} & \colhead{DIT}  & \colhead{Survey}\\
\colhead{\#} & \colhead{(J2000)} & \colhead{(J2000)} & \colhead{} & \colhead{} & \colhead{} & \colhead{group} & \colhead{(YYYY-MM-DD)} & \colhead{(min)} & \colhead{Rot. ($^{\circ}$)} & \colhead{} & \colhead{(sec)} & \colhead{ID}}
%\colnumbers
\startdata
%50520 & 11 & 6.23 & A1V & LCC & & & & K\\
%53524 & A1 & 6.76 & A8III & LCC & & & &\\
%54231 & A2 & 6.75 & A0V & LCC & & & & H\\
%55188 & 41 & 7.43 & A2V & LCC & & & & K\\
%55899 & A3 & 7.07 & A0V & LCC   & & & & &  \\
%56354 & 6 & 5.78 & A9V & LCC & & & & K\\
%56993 & 44 & 7.38 & A0V & LCC & & & & K\\
%57809 & 47 & 6.61 & A0V & LCC & & & & K\\
%58465 & A5 & 6.32 & A2V & LCC & & & & H\\ 
%59282 & A6 & 7.00 & A3V & LCC   & & & & &  \\
%59397 & 51 & 7.01 & A2V & LCC & & & & K\\
44&14 32 57.1 &-42 24 20& 71140 & 7.13& A7/A8IV  &      UCL   & 2015-04-29 & 42.7 & 34.8 & $K12$ & 8.00 & S12\\
& &&  & &  &          & 2019-06-29 & 29.9 & 25.3 & $K12$ & 8.00 & S12\_2\\
45&14 34 33.4 &-46 18 17& 71271 & 7.57& A0V       &     UCL   & 2017-05-29 & 65.1 & 48.1 & $H$ & 32.0 & SA21\\
46&14 40 19.3 &-45 47 38 &71727 & 6.89& A0p     &       UCL   & 2015-04-06 & 21.1 & 24.0 & $K12$ & 8.00 & S3\\
& &&  & &  &          & 2019-06-27 & 29.6 & 21.5 & $K12$ & 8.00 & S3\_2\\
47&14 45 20.6 &-36 08 52& 72140 & 7.09& A1IV/V  &       UCL   & 2016-04-06 & 25.6 & 30.7 & $K12$ & 16.0 & S34\\
& &&  & &  &          & 2019-05-26 & 29.9 & 35.5 & $K12$ & 8.00 & S34\_2\\
48&14 45 57.6 &-44 52 03& 72192 & 6.71& A0V      &      UCL   & 2015-06-11 & 25.6 & 17.3 & $K12$ & 16.0 & S21\\
49&14 50 58.7 &-42 49 21& 72627 & 6.53& A2V       &     UCL   & 2016-04-23 & 24.0 & 19.6 & $K12$ & 16.0 & S36\\
& &&  & &  &          & 2019-05-26 & 29.9 & 23.4 & $K12$ & 8.00 & S36\_2\\
50&14 54 25.3 &-34 08 34& 72940 & 6.85& A1V        &    UCL   & 2015-06-12 & 25.6 & 36.8 & $K12$ & 16.0 & S22\\
& &&  & &  &          & 2016-03-06 & 33.6 & 38.5 & $K12$ & 32.0 & S22\_2\\
& &&  & &  &          & 2016-03-17 & 29.9 & 36.6 & $K12$ & 32.0 & S22\_3\\
& &&  & &  &          & 2016-05-07 & 29.9 & 39.6 & $K12$ & 32.0 & S22\_4\\
& &&  & &  &          & 2017-05-31 & 27.7 & 23.1 & $Y23$ & 16.0 & S22\_5\\
& &&  & &  &          & 2017-07-22 & 27.7 & 3.89 & $H23$ & 16.0 & S22\_6\\
& &&  & &  &          & 2017-07-23 & 26.1 & 5.59 & $K12$ & 32.0 & S22\_7\\
& &&  & &  &          & 2017-07-23 & 24.5 & 3.78 & $Ks$ & 32.0 & S22\_8\\
& &&  & &  &          & 2017-07-24 & 27.2 & 10.7 & $J23$ & 16.0 & S22\_9\\
& &&  & &  &          & 2017-07-28 & 26.9 & 10.2 & $H23$ & 16.0 & S22\_10\\
& &&  & &  &          & 2018-03-14 & 27.5 & 38.9 & $H23$ & 16.0 & S22\_11\\
& &&  & &  &          & 2018-07-14 & 34.1 & 38.0 & $K12$ & 16.0 & S22\_12\\
51&14 54 54.5 &-36 25 49& 72984 & 7.05& A0/A1V     & UCL  & 2018-03-16 & 40.0 & 52.4 & $K12$ & 16.0 & S61\\
52&14 56 54.5 &-35 41 44& 73145 & 7.54& A2IV    &       UCL   & 2015-05-14 & 67.2 & 72.5 & $H23$ & 64.0 & SA22\\
53&14 59 54.6 &-46 14 53& 73393 & 7.21& A0V      &      UCL   & 2015-07-17 & 25.6 & 14.2 & $K12$ & 16.0 & S29\\
& &&  & &  &          & 2019-05-10 & 29.9 & 21.2 & $K12$ & 8.00 & S29\_2\\
54&15 06 33.2 &-30 55 07& 73937 & 6.05& Ap        &     UCL   & 2018-03-16 & 32.0 & 21.9 & $K12$ & 16.0 & S62\\
55&15 17 10.7 &-34 34 37& 74797 & 7.55& A2IV       &    UCL   & 2015-04-29 & 23.5 & 33.9 & $K12$ & 8.00 & S13\\
& &&  & &  &          & 2019-06-27 & 22.3 & 30.3 & $K12$ & 8.00 & S13\_2\\
56&15 19 22.3 &-34 01 57& 74985 & 7.53& A0V         &   UCL   & 2017-05-15 & 31.2 & 45.1 & $K12$ & 16.0 & S49\\
57&15 20 13.4 &-34 55 32& 75056 & 7.31& A2V          &  UCL    & 2015-06-19 & 21.3 & 13.8 & $K12$ & 16.0 & S25\\
& &&  & &  &          & 2019-06-29 & 28.5 & 41.6 & $K12$ & 8.00 & S25\_2\\
58&15 20 31.4 &-28 17 14& 75077 & 6.97& A1V     &       UCL  & 2016-04-03 & 16.4 & 25.7 & $H23$ & 4.00 & SA23\\
59&15 21 30.1 &-38 13 07& 75151 & 6.65& A+...    &      UCL   & 2015-07-18 & 25.6 & 18.7 & $K12$ & 16.0 & S31\\
%&15 25 06.4 &-38 10 09& 75476 & 6.88& A1/A2V    &     UCL   & & & & &  SA24 \\
60&15 25 30.2 &-36 11 58& 75509 & 7.40& A2V        &    UCL    & 2018-04-10 & 76.8 & 75.6 & $K12$ & 96.0 & SA25\\
61&15 30 48.4 &-45 25 28& 75957 & 7.24& A0V         &   UCL   & 2015-07-12 & 41.9 & 27.8 & $K12$ & 16.0 & S28\\
& &&  & &  &          & 2019-05-01 & 29.9 & 21.7 & $K12$ & 8.00 & S28\_2\\
%&15 35 16.1 &-25 44 03& 76310 & 7.35& A0V          &   US   & & & & &  \\
%&15 45 06.4 &-35 06 07& 77150 & 7.28& A2V           & UCL   & & & & &  \\
62&15 46 51.6 &-36 56 13& 77295 & 7.64& A2IV/V  &       UCL   & 2017-05-15 & 32.0 & 32.2 & $K12$ & 16.0 & S50\\
63&15 48 52.1 &-29 29 00& 77457 & 7.33& A7IV     &       US   & 2015-06-09 & 25.6 & 12.0 & $K12$ & 16.0 & S18\\
& &&  & &  &          & 2017-05-15 & 32.0 & 14.7 & $K12$ & 16.0 & S18\_2\\
64&15 56 47.9 &-23 11 03& 78099 & 7.35& A0V       &      US   & 2015-05-09 & 27.7 & 8.19 & $H23$ & 64.0 & SA26\\
65&15 57 59.3 &-31 43 44& 78196 & 7.08& A0V        &     US   & 2015-06-03 & 57.6 & 84.6 & $H23$ & 64.0 & SA27\\
66&16 01 26.6 &-25 11 55& 78494 & 7.11& A2m...      &    US    & 2015-05-18 & 22.1 & 2.34 & $K12$ & 8.00 & S16\\
67&16 01 58.9 &-37 32 04& 78533 & 6.99& Ap           &  UCL     & 2016-07-14 & 24.3 & 24.8 & $K12$ & 16.0 & S45\\
& &&  & &  &          & 2019-05-12 & 36.3 & 32.1 & $K12$ & 8.00 & S45\_2\\
68&16 02 04.8 &-36 44 38& 78541 & 6.99& A0V           & UCL   & 2015-05-19 & 23.5 & 12.5 & $K12$ & 8.00 & S17\\
69&16 04 44.5 &-39 26 05& 78756 & 7.16& Ap             &UCL    & 2016-04-23 & 25.1 & 18.4 & $K12$ & 16.0 & S37\\
& &&  & &  &          & 2016-08-07 & 22.7 & 20.9 & $K12$ & 16.0 & S37\_2\\
& &&  & &  &          & 2017-05-29 & 31.5 & 28.3 & $K12$ & 16.0 & S37\_3\\
70&16 05 43.4 &-21 50 20& 78847 & 7.32& A0V             &US   & 2015-06-17 & 25.6 & 2.59 & $K12$ & 16.0 & S24\\
71&16 05 46.3 &-39 50 36& 78853 & 7.50& A5V     &       UCL   & 2016-04-23 & 19.5 & 23.9 & $K12$ & 16.0 & S38\\
72&16 07 29.9 &-23 57 02& 78996 & 7.46& A9V      &       US  & 2015-06-19 & 25.6 & 0.558 & $K12$ & 16.0 & S26\\
73&16 11 52.7 &-22 32 42& 79366 & 7.47& A3V       &      US   & 2017-07-28 & 29.9 & 2.65 & $K12$ & 16.0 & S59\\
74&16 18 05.5 &-31 39 06& 79860 & 7.88& A0V        &     US   & 2015-07-31 & 18.7 & 37.6 & $K12$ & 16.0 & S33\\
75&16 18 16.2 &-28 02 30& 79878 & 7.06& A0V         &    US  & 2016-07-02 & 40.4 & 17.9 & $H23$ & 8.00 & SA28\\
76&16 20 04.0 &-20 02 42& 80019 & 7.08& A0V          &   US   & 2015-04-21 & 23.5 & 355. & $K12$ & 8.00 & S9\\
& &&  & &  &          & 2019-05-31 & 29.9 & 36.2 & $K12$ & 8.00 & S9\_2\\
77&16 20 28.1 &-21 30 32& 80059 & 7.44& A7III/IV      &  US  & 2017-07-25 & 32.0 & 4.21 & $K12$ & 16.0 & S57\\
78&16 23 56.7 &-33 11 58& 80324 & 7.33& A0V        & US   & 2018-06-18 & 59.2 & 53.1 & $H23$ & 96.0 & SA29\\
79&16 25 35.1 &-23 24 19& 80474 & 5.76& A               &US  & 2017-07-27 & 32.0 & 0.252 & $K12$ & 16.0 & S58\\
80&16 27 14.6 &-39 49 22& 80591 & 7.82& A5V         &   UCL   & 2015-04-05 & 20.4 & 30.0 & $K12$ & 8.00 & S2\\
& &&  & &  &          & 2017-05-29 & 26.4 & 28.8 & $K12$ & 16.0 & S2\_2\\
81&16 29 54.6 &-24 58 46& 80799 & 7.46& A2V          &   US  & 2016-08-07 & 25.3 & 2.09 & $K12$ & 16.0 & S46

\enddata
\tablecomments{Continued from Table 2. Survey IDs were assigned in numerical order as observations were completed, except for those obtained from the archive (those beginning with 'SA'), which were assigned in order of increasing R.A. (Right Ascension). The other abbreviations in the column headers are Dec. (Declination), SpT (Spectral Type), and DIT (Detector Integration Time).}
%\tablenotetext{a}{\cite{Rameau2013}}
%\tablenotetext{a}{\cite{Chauvin2017}}
%\tablecomments{...}
\end{deluxetable*}

\begin{deluxetable*}{ccccccccccccc}
%\startlongtable

\tablecaption{Targets $-$ Sco-Cen A-type Stars (Continued)}%\tablenotemark{a} \label{tab:table}}
\tablehead{
\colhead{Target} & \colhead{R.A.} & \colhead{Dec.} & \colhead{HIP \#} & \colhead{$K_{s}$} & \colhead{SpT} & \colhead{Sub-} & \colhead{Date Obs.} & \colhead{Exp. Time} & \colhead{Field} & \colhead{Filter} & \colhead{DIT}  & \colhead{Survey}\\
\colhead{\#} & \colhead{(J2000)} & \colhead{(J2000)} & \colhead{} & \colhead{} & \colhead{} & \colhead{group} & \colhead{(YYYY-MM-DD)} & \colhead{(min)} & \colhead{Rot. ($^{\circ}$)} & \colhead{} & \colhead{(sec)} & \colhead{ID}}
%\colnumbers
\startdata

82&16 31 11.7 &-38 22 59& 80897 & 7.78& A0V           & UCL    & 2015-07-19 & 23.5 & 19.5 & $K12$ & 16.0 & S32\\
& &&  & &  &          & 2017-05-28 & 32.0 & 29.7 & $K12$ & 16.0 & S32\_2\\
83&16 34 10.5 &-38 23 25& 81136 & 5.21& A7/A8    &  UCL   & 2016-04-23 & 26.3 & 28.4 & $K12$ & 8.00 & S39\\
& &&  & &  &          & 2019-05-01 & 29.9 & 32.3 & $K12$ & 8.00 & S39\_2\\
84&16 41 52.0 &-40 27 58& 81751 & 8.29& A9V           & UCL   & 2015-04-29 & 23.5 & 22.0 & $K12$ & 8.00 & S14\\
& &&  & &  &          & 2017-06-22 & 31.7 & 27.2 & $K12$ & 16.0 & S14\_2\\
85&16 44 21.1 &-44 22 43& 81949 & 7.32& A3V          &  UCL    & 2016-04-23 & 25.1 & 15.7 & $K12$ & 16.0 & S40\\
%&16 50 10.7 &-26 44 33& 82397 & 7.28& A3V          &   US   & & & & &  \\
86&16 52 31.9 &-42 59 30& 82560 & 6.58& A0V          &  UCL  & 2016-05-20 & 25.6 & 17.5 & $K12$ & 16.0 & S43\\
& &&  & &  &          & 2019-04-28 & 29.9 & 24.1 & $K12$ & 8.00 & S43\_2\\
87&17 03 22.2 &-40 05 16& 83457 & 6.49& A9V         &   UCL    & 2015-04-15 & 23.2 & 19.4 & $K12$ & 8.00 & S8\\
& &&  & &  &          & 2016-04-23 & 21.3 & 11.4 & $K12$ & 16.0 & S8\_2\\
88&17 06 20.2 &-37 13 39& 83693 & 5.69& A2IV    &       UCL   & 2015-04-23 & 21.9 & 28.0 & $K12$ & 8.00 & S10\\
& &&  & &  &          & 2016-08-11 & 25.5 & 25.8 & $K12$ & 8.00 & S10\_2\\
& &&  & &  &          & 2017-05-29 & 30.1 & 32.1 & $K12$ & 16.0 & S10\_3

\enddata
\tablecomments{Continued from Table 3. Survey IDs were assigned in numerical order as observations were completed, except for those obtained from the archive (those beginning with 'SA'), which were assigned in order of increasing R.A. (Right Ascension). The other abbreviations in the column headers are Dec. (Declination), SpT (Spectral Type), and DIT (Detector Integration Time).}
%\tablenotetext{a}{\cite{Rameau2013}}
%\tablenotetext{a}{\cite{Chauvin2017}}
%\tablecomments{...}
\end{deluxetable*}

%\appendix

%\section{...}

\end{document}